\documentclass[12pt]{article}
\usepackage[left=2.5cm,right=2.5cm,top=2cm,bottom=2cm]{geometry} 
\usepackage{floatrow}
\usepackage{tabularray}
\usepackage[table]{xcolor}
\usepackage{listings}
\usepackage[numbers,sort&compress]{natbib}
\usepackage{caption}
\usepackage{subcaption}
\DeclareMathAlphabet\mathbfcal{OMS}{cmsy}{b}{n}
\usepackage{amsmath} 

\usepackage{graphicx,caption}
\usepackage[none]{hyphenat}
\usepackage{mathtools}
\usepackage{tkz-euclide}
\usepackage[english]{babel}
\usepackage{epsfig}
\usepackage{pgfplots}
\usepackage{color,soul}    
\usepackage[normalem]{ulem}
\definecolor{dkgreen}{rgb}{0,0.6,0}
\definecolor{gray}{rgb}{0.5,0.5,0.5}
\definecolor{mauve}{rgb}{0.58,0,0.82}
\definecolor{green}{HTML}{B9E69B}
\definecolor{orange}{HTML}{F8DC7A}
\definecolor{blue}{HTML}{0095FF}
\usepackage{graphicx,calc}
\newlength\myheight
\newlength\mydepth
\settototalheight\myheight{Xygp}

\DeclareRobustCommand{\mychar}{%
  \begingroup\normalfont
  \includegraphics[height=\myheight]{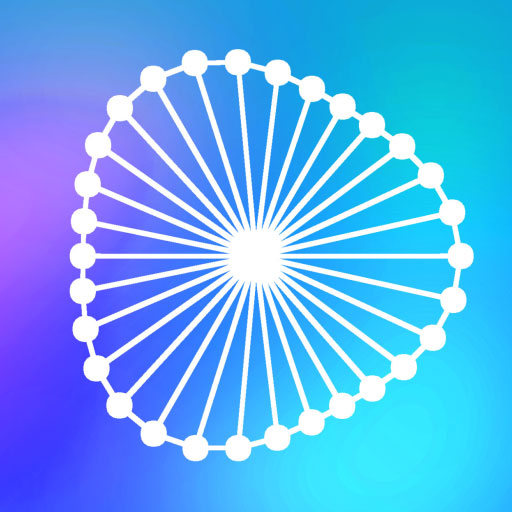}%
  \endgroup
}

\usepackage{xcolor,colortbl}
\usepackage{lscape}
\usepackage[utf8]{inputenc}
\usepackage{amsmath,bm}
\usepackage{amssymb,amsbsy}
\usepackage{amsthm}
\usepackage{authblk}
\usepackage{tikz}
\usetikzlibrary{arrows.meta,calc,patterns,decorations.pathmorphing,decorations.markings,positioning,shapes.geometric, quotes}
\usepackage{graphicx}
\usepackage{comment}
\usepackage{subcaption}
\usepackage[export]{adjustbox}
\usetikzlibrary{arrows}
\usepackage{tikz-3dplot}
\usepackage{enumitem}
\usepackage[toc,page]{appendix}

\usepackage{microtype}

\title{A mechanical analogue of electromagnetic induction for waves in a chiral elastic structure}
\author{Finn J.P. Allison\thanks{Corresponding author: F.Allison2@liverpool.ac.uk}, \"Ozg\"ur Selsil, Stewart G. Haslinger \& Alexander B. Movchan}
\affil{Department of Mathematical Sciences, University of Liverpool}

\numberwithin{equation}{section}

\begin{document}
\microtypesetup{activate=true}

\maketitle

\begin{abstract} 
Classical Faraday's law on electromagnetic induction states that a change of magnetic field through a coil wire induces a current in the wire. A mechanical analogue of the Lorentz force, induced by a magnetic field on an electric charge, is the gyroscopic force. Here, we demonstrate a mechanical analogy with a chiral elastic waveguide subjected to gyroscopic forcing. We study waves in an infinite mass-spring `gyrocore helix', which consists of a helix and a central line (gyroscopic elastic core). The helicoidal geometric chirality is considered in conjunction with a physical chirality, induced by  gyroscopic forces. It is shown that the interplay between these two chiral inputs leads to the breaking of symmetry of the associated dispersion diagram, resulting in a unidirectional waveform with the direction of propagation being tunable through the gyricity.
\end{abstract}

\section{Introduction}

\subsection{Bibliographical remarks}


The formal mathematical analogy between the Lorentz force acting on an electric charge, moving in a magnetic field, and the mechanical gyroscopic force acting on a moving body is well established in the classical books by Gray \cite{gray1918treatise} (chapters IX and XX) and Webster \cite{webster1912dynamics} (chapters VI and VII). \\

In mechanics, the notion of {\em physical chirality} is in the dynamic response of a gyroscopic system, which exhibits a rotational motion of a certain orientation. Classical examples of physically chiral systems include Laplace's theory of tidal waves \cite{laplace1823traite} and Lighthill's theory of equatorial monsoons \cite{lighthill1969dynamic}. In both cases, the governing equations of shallow water waves incorporate the Coriolis force orthogonal to the velocity vector. The presence of vortices in the atmosphere, due to the Coriolis force, is also well known: around a region of depression in the Northern hemisphere the air moves along the counterclockwise vortex, whereas in the Southern hemisphere the vortex around the region of depression is clockwise.    \\
 
The classical {\em geometric chirality} is in the notion of ``handedness'', which is well known, and it is applicable to both static and dynamic structures observed in nature. According to Lord Kelvin \cite{kelvin1894molecular}, any geometric figure, or group of points,  is chiral if its image in a plane mirror cannot be brought to coincide with itself. For elastic waves propagating in geometrically chiral helical springs, the coupling between longitudinal displacements and rotations was analysed in \cite{wittrick1966elastic} and important asymptotic analysis for helical waveguides was published in \cite{sorokin2009linear}. \\

When several gyroscopic elements are assembled into a cluster within an elastic structure, the mutual interaction between gyroscopic constituents lead to an interesting dynamic response.  
Physically chiral periodic elastic lattice systems, subjected to gyroscopic forces, have been extensively studied  for 
  two-dimensional doubly periodic structures \cite{carta2019wave, garau2018interfacial, carta2020one, carta2020chiral, carta2017deflecting} and one-dimensional coupled chains \cite{jones2020two}. Further studies include analysis of chiral waveforms in periodic gyroscopic systems, subjected to gravity \cite{kandiah2023effect, kandiah2024controlling}. In \cite{carta2019wave}, a triangular lattice with embedded gyroscopic spinners was shown to induce so-called vortex waves;  strong dynamic anisotropy for Bloch-Floquet waves was observed, and their dispersion was affected by the physical chirality of the system. Physically chiral systems, incorporating gyroscopic dynamic response, may be used in unidirectional wave steering to create topological insulators \cite{ni2015topologically,susstrunk2015observation,garau2018interfacial}. The papers \cite{carta2020one,carta2020chiral}, include analysis of the dispersion properties of chiral periodic gyroscopic elastic systems, where the gyroscopic action is analogous to the action of a unidirectional magnetic field in photonic crystal structures. Novel geometrically chiral elastic systems, which couple dilatational and rotational vibration modes, have been studied in \cite{bigoni2013elastic, tallarico2017tilted, frenzel2017three}; these systems did not include gyroscopic forces.   Here, we consider a new chiral object, represented by a three-dimensional helix-like elastic lattice, supported by a central gyroscopic core aligned with the axis of the helix. Such a structure incorporates both chiral features: geometric chirality of the helix and the physical chirality of the central gyroscopic core. \\

\subsection{Overview of the paper}

Faraday’s law of electromagnetic induction states that a changing magnetic field induces a voltage in a coil of wire. Indeed, the induced current $I$ may also be given in terms of the rate of change of the magnetic flux $\phi$ as 
\begin{equation}
    I = -\frac{1}{R} \frac{\text{d} \phi}{\text{d} t},
    \label{faraday}
\end{equation}
where  $R$ represents the resistance of the  coil. 
The coil's geometric chirality, or handedness, affects the orientation of the induced magnetic field relative to the changing magnetic flux. Consequently, for a given change in flux, the direction of the induced current will differ between right-handed and left-handed coils. \\

\begin{figure}[H]
    \centering
\includegraphics[scale=0.2]{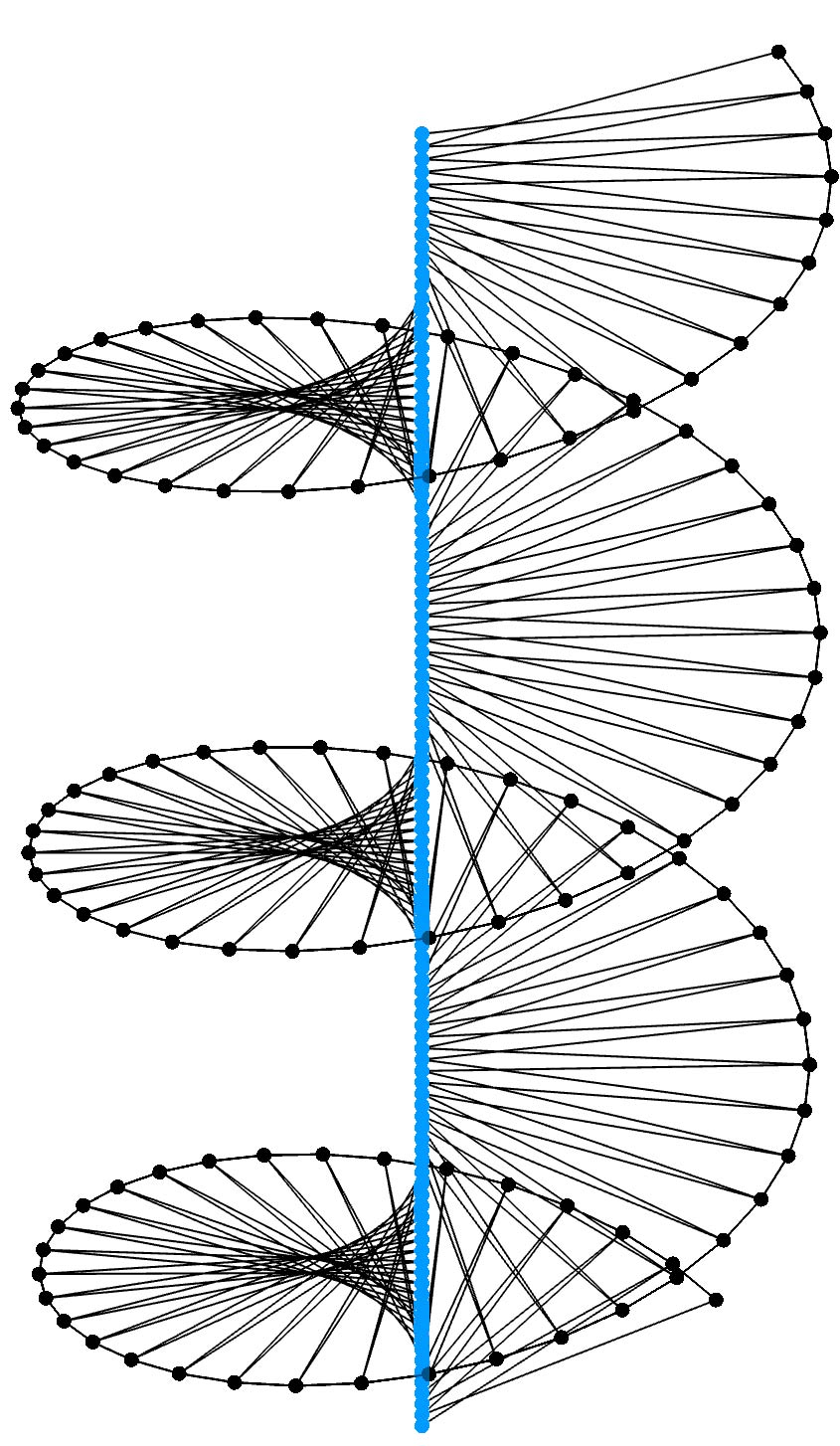}
    \caption{A portion of the infinite `gyrocore helix', where masses with gyroscopic spinners attached are depicted in blue. Use the Eigenglass application to view an enhanced three-dimensional representation. \mychar{}}
    \label{sampleconfig}
\end{figure}

\begin{figure}
    \centering
\begin{tikzpicture}
\node[]
    {\includegraphics[scale=0.5]{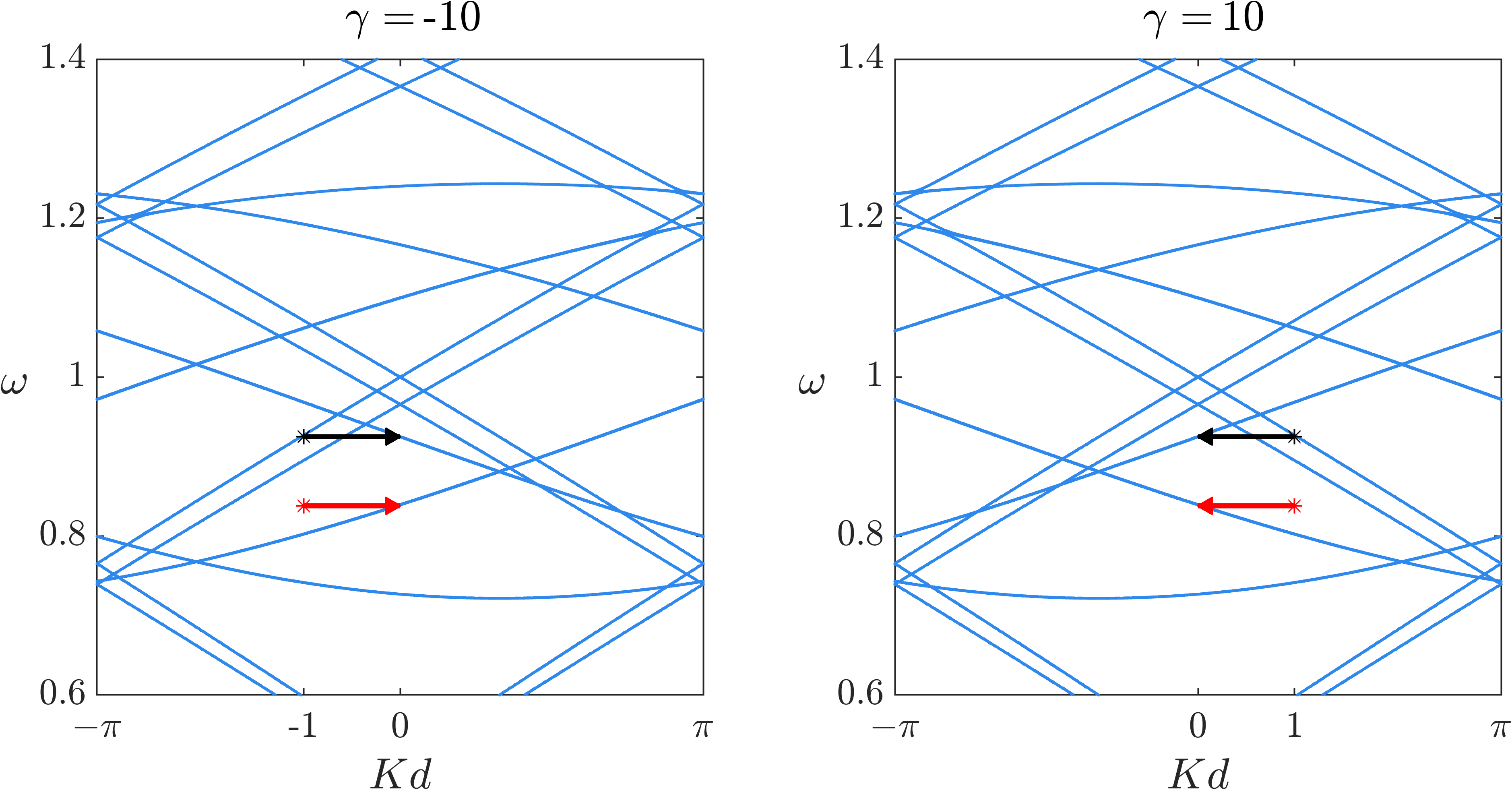}};
\node[black] at (-3.8,-4.8) {$(a)$};
\node[black] at (4.7,-4.8) {$(b)$};
\end{tikzpicture}
  \caption{A close-up of the dispersion diagram for a right-handed gyrocore helix for (a) $\gamma = -10$ and (b) $\gamma = 10$. Note a right- and left-ward shift of the dispersion branches, respectively. The arrows indicate the higher (black) and lower (red) split eigenfrequencies at $Kd=\mp1$ which are shifted to $Kd=0$. Use the Eigenglass application (Supplementary Material) to see how the amount of deviation from the original dispersion curves is dependent on the spinner constant $\gamma$. \mychar{}}
    \label{disp_group_vel}
\end{figure}

In this article, we investigate an analogous effect for a mechanical system. Considering a `gyrocore helix', that is a discrete, mass-spring helix with gyroscopes attached to the masses on its central line (see Figure \ref{sampleconfig} - note that blue masses represent those with gyroscopic spinners attached), we demonstrate that, for a fixed frequency, the direction of the group velocity $v_g$ is determined by the gyricity $\mathcal{G}$ in relation to the handedness of the helix. This is exemplified in the dispersion diagram as a left or right shift of specific dispersion branches and a splitting of the eigenfrequencies at wavenumber $K=0$
\begin{equation}
    v_g = \pm\frac{\omega_{*}^{\pm}}{\mathcal{G}}.
    \label{groupvelocity}
\end{equation}
 Here, $\omega_*$ is a constant frequency, with the superscripts $^\pm$ denoting the higher and lower split frequency values, respectively. In Figure \ref{disp_group_vel}, it is apparent that the degree of shift is dependent on the magnitude of the spinner constant $\gamma$, with the shift direction specified by its sign. For gyroscopes which spin against the handedness of the helix, we observe a rightward shift of the dispersion branches with the lower frequency eigenmodes possessing a positive group velocity (Figure \ref{disp_group_vel}(a)); for the same frequency, we note a negative group velocity when the vorticity of the gyroscopes is inverted (Figure \ref{disp_group_vel}(b)). The black and red arrow heads indicate the higher and lower split frequencies $\omega^{\pm}$, respectively, with their length describing the degree of shift from $Kd=\mp1$, where \(Kd\) is the normalised wavenumber with \(d\) being the height (pitch) of the macrocell (helix). \\
\subsubsection{Vibrational Modes}
The aforementioned shift of specific dispersion branches gives rise to vibrational modes which possess a non-zero group velocity with zero wavenumber. The three-dimensional visualisations of such modes (given in the Supplementary Material) depict oscillations which are absent of spatial variation along the $z$-axis, yet maintain a positive or negative group velocity. The branches in question pertain to shear vibrations of the central line masses which, for the frequencies indicated in Figure \ref{disp_group_vel}, organise to form an internal helix - mimicking the structure it is entwined with. These `helical core' modes are characterised by their spin direction and geometric handedness, the former being dependent on the split frequency under consideration. For the lower frequency split modes, (see red arrows in Figure \ref{disp_group_vel}), the spin direction of the helical core opposes that of the gyroscopic spinners. Whilst the higher frequency modes (see black arrows in Figure \ref{disp_group_vel}) spin in agreement with the direction of the gyroscopes. We conclude that the vorticity of the gyroscopes determines the group velocity direction for a fixed frequency which subsequently determines the spin direction of these helical core vibrational modes.

\subsubsection{Physical chirality in the governing equations}

Physical chirality pertains to the behaviour of gyroscopic systems in motion. 
It encompasses asymmetries that arise from gyroscopic forces and dynamic interactions, rather than being solely dependent on their static geometric arrangement. 
Formally, the gyroscopic forces are represented by the last term containing the coupling matrix \(\textbf{R}\) and the velocity components in equation (\ref{central_line_gov_eqn}) and first two equations of (\ref{remaining_gov_eqns}).\\


\subsubsection{Analogy}

\begin{table}[h!]
\begin{tblr}{
  hlines,
  vlines,
  row{1} = {bg=gray,fg=white},
  rows = {ht=1cm},
  columns = {halign=c},
  colspec = {Q[bg=gray,fg=white] XXX},
  } 
  & Geometric chirality & Physical chirality & Affected quantity \\
  Electromagnetic & Coiled wire  & Change in magnetic flux  & Electric current \\
  Mechanical  & Mass-spring helix & Gyricity  & Group velocity \\
\end{tblr}
\caption{A table of compared quantities for the mechanical analogue of electromagnetic induction.}
\label{go compare}
\end{table}

Table \ref{go compare} refers to geometric and physical chirality of two helical systems,  the induction coil and the mass-spring gyrocore helix. The helicoidal structure of both systems represents the geometric chirality, while the physical chirality is induced by the magnetic field in the case of induction coil and by the gyroscopic forces for the corresponding mechanical system. Formulae (\ref{faraday}) and (\ref{groupvelocity}) give the electric current in the coil wire and the group velocity of the elastic wave, respectively. \\

\subsubsection{Eigenglass application}
In order to portray the three-dimensional vibrational modes, we designed a novel tool, a mobile app, that would not only provide important context and exposition to the analysis, but enhance the reading experience. For this reason, we highly encourage the reader to download the Eigenglass application, available on App store and Google Play, and take time to explore the dynamic dispersion diagram animations and three-dimensional augmentations hidden behind each figure with \mychar{}. For those unable to download the application, all visualisations are also available as part of the Supplementary Material (Section 2).
\\

\subsubsection{Structure of the paper}
We convolve the helix-like structure with a physical chirality in Section \ref{prob_formulation} by attaching a series of gyroscopic spinners to its central line masses. The resulting structure, referred to as a gyrocore helix, possesses both physical and geometric chiralities, determined by the vorticity of the spinners and the handedness of the helix, respectively. In Section \ref{solutionsec}, we present the solution method and discuss the low-frequency behaviour of the gyrocore helix. 
Section \ref{Results} is dedicated to the key results, and concluding remarks are made in Section \ref{Conclusion} .

\section{Problem formulation}
\label{prob_formulation}
\subsection{Helix-like structure}
We now formulate the discrete, geometrically chiral, helix-like structure. We use this terminology not only because the structure is discrete but also because it consists of a helical rim as well as a central line, the latter having been introduced to add structural integrity. \\
In the generating unit cell $\Omega^{(0)}$, the building block of the construction, both the helix and central line have $M$ masses attached to each other via massless springs, denoted by $m^{(0)}_{h,k}$ and $m^{(0)}_{c,k}, \, k=1,\dots,M$, where subscripts $h$ and $c$ stand for helix and central, respectively. In addition, each mass on the helix is attached to two masses on the central line, more precisely $m^{(0)}_{h,k}$ is connected to both $m^{(0)}_{c,k\pm1}$. An infinite three-dimensional helix-like structure is thus constructed by stacking $\Omega^{(0)}$ vertically along the positive and negative $z$-axis; we enumerate the masses in the proceeding/preceding unit cells as $m^{(j)} \in \Omega^{(j)},\, j \in \mathbb{Z}_{+/-}$, respectively. We present a sample configuration in Figure \ref{sampleconfig}, with springs depicted as straight lines for visual simplicity. \\

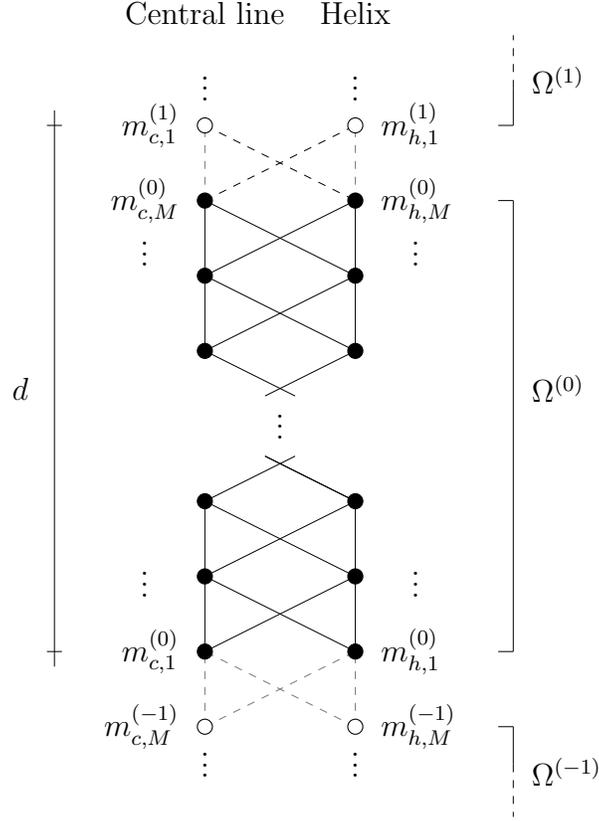
\begin{figure}[h!]
    \centering
    
    \begin{tikzpicture}

 \foreach \n in {1,2,5,6}
    \draw[black,very thin] (0,\n) -- (2,\n+1);
    \draw[black,very thin, dashed] (0,7) -- (2,8);
    
   \draw[black!50,dashed] (0,0) -- (0,1);
      \draw[black!50,dashed] (0,7) -- (0,8);
         \draw[black!50,dashed] (2,7) -- (2,8);
    \draw[black!50,dashed] (0,0) -- (2,1);
     \draw[black!50,dashed] (2,0) -- (0,1);
      \draw[black!50,dashed] (2,0) -- (2,1);
      
    
     \foreach \n in {2,3,6,7}
    \draw[black,very thin] (0,\n) -- (2,\n-1);
    \draw[black,very thin, dashed] (0,8) -- (2,7);

    \draw[black,very thin] (0,1) -- (0,3);
    \draw[black,very thin] (0,5) -- (0,7);
    \draw[black,very thin] (2,1) -- (2,3);
    \draw[black,very thin] (2,5) -- (2,7);
    \draw[black,very thin] (2,3) -- (0.8,3.6);
    \draw[black,very thin] (0,3) -- (1.2,3.6);
    \draw[black,very thin] (2,3) -- (0.8,3.6);
    \draw[black,very thin] (0,5) -- (1.2,4.4);
    \draw[black,very thin] (2,5) -- (0.8,4.4);

 \foreach \n in {1,2,3,5,6,7}
    \draw[black,fill] (0,\n) circle [radius=1mm];
    
    \foreach \n in {1,2,3,5,6,7}
    \draw[black,fill] (2,\n) circle [radius=1mm];

\draw[black,fill=white] (0,0) circle [radius=1mm];
\draw[black,fill=white] (0,8) circle [radius=1mm];
\draw[black,fill=white] (2,0) circle [radius=1mm];
\draw[black,fill=white] (2,8) circle [radius=1mm];

\node[left=0.2cm] at (0,0) {$m_{c,M}^{(-1)}$};
\node[left=0.2cm] at (0,8) {$m_{c,1}^{(1)}$};
\node[right=0.2cm] at (2,0) {$m_{h,M}^{(-1)}$};
\node[right=0.2cm] at (2,8) {$m_{h,1}^{(1)}$};

\node[left=0.2cm] at (-0.4,2) {$\vdots$};
\node[right=0.2cm] at (2.4,2) {$\vdots$};
\node[left=0.2cm] at (-0.4,6.4) {$\vdots$};
\node[right=0.2cm] at (2.4,6.4) {$\vdots$};
\node at (1,4.1) {$\vdots$};

\node[left=0.2cm] at (0,1) {$m_{c,1}^{(0)}$};
\node[left=0.2cm] at (0,7) {$m_{c,M}^{(0)}$};
\node[right=0.2cm] at (2,1) {$m_{h,1}^{(0)}$};
\node[right=0.2cm] at (2,7) {$m_{h,M}^{(0)}$};

\node at (0,9.5) {Central line};
\node at (2,9.5) {Helix};

\draw[ black, very thin] (-2,0.8)--(-2,8.2);
\draw[ black, very thin] (-2.1,1)--(-1.9,1);
\draw[ black, very thin] (-2.1,8)--(-1.9,8);
\node[left=0.2cm] at (-2,4.5) {$d$};

\draw[ black, very thin] (3.9,1)--(4.1,1);
\draw[ black, very thin] (3.9,7)--(4.1,7);
\draw[ black, very thin] (4.1,1)--(4.1,7);
\node[right=0.2cm] at (4,4.5) {$\Omega^{(0)}$};
\draw[ black, very thin] (3.9,8)--(4.1,8);
\draw[ black, very thin] (4.1,8)--(4.1,8.5);
\draw[ black, very thin, dashed] (4.1,8.5)--(4.1,9.2);
\node[right=0.2cm] at (4,8.6) {$\Omega^{(1)}$};
\draw[ black, very thin] (3.9,0)--(4.1,0);
\draw[ black, very thin] (4.1,0)--(4.1,-0.5);
\draw[ black, very thin, dashed] (4.1,-0.5)--(4.1,-1.2);
\node[right=0.2cm] at (4,-0.6) {$\Omega^{(-1)}$};

\node at (0,-0.4) {$\vdots$};
\node at (2,-0.4) {$\vdots$};
\node at (0,8.6) {$\vdots$};
\node at (2,8.6) {$\vdots$};

\end{tikzpicture}
    \caption{An unwound two-dimensional representation of the infinite (three-dimensional) helix-like structure. 
    The dashed lines indicate the connections between the generating unit cell $\Omega^{(0)}$ and its repeated copies below and above, namely the neighbouring unit cells $\Omega^{(-1)}$ and $\Omega^{(1)}$. \mychar{}
}
    \label{2D_helix-like_config}
\end{figure}
\begin{figure}[h!]   
\begin{center}
\begin{tikzpicture}

\draw[black, very thin] (0,-4)--(0,-3);
\draw[black, very thin] (0,-4)--(2,-3);
\draw[black, very thin, dashed] (0,-4)--(0,-5);
\draw[black, very thin,dashed] (0,-4)--(2,-5);

\draw[black,fill] (0,-4) circle [radius=1mm];
\draw[black,fill] (0,-3) circle [radius=1mm];
\draw[black,fill] (2,-3) circle [radius=1mm];

\draw[black,fill=white] (0,-5) circle [radius=1mm];
\draw[black,fill=white] (2,-5) circle [radius=1mm];

\node[left=0.2cm] at (0,-4) {$m_{c,1}^{(0)}$};
\node[left=0.2cm] at (0,-5) {$m_{c,M}^{(-1)}$};
\node[right=0.2cm] at (2,-5) {$m_{h,M}^{(-1)}$};
\node at (1,-6) {(c)};

\draw[black, very thin,dashed] (8,0)--(6,1);
\draw[black, very thin,dashed] (8,0)--(8,1);
\draw[black, very thin] (8,0)--(6,-1);
\draw[black, very thin] (8,0)--(8,-1);

\draw[black,fill] (8,0) circle [radius=1mm];
\draw[black,fill=white] (6,1) circle [radius=1mm];
\draw[black,fill=white] (8,1) circle [radius=1mm];

\draw[black,fill] (6,-1) circle [radius=1mm];
\draw[black,fill] (8,-1) circle [radius=1mm];

\node[right=0.2cm] at (8,0) {$m_{h,M}^{(0)}$};
\node[left=0.2cm] at (6,1) {$m_{c,1}^{(1)}$};
\node[right=0.2cm] at (8,1) {$m_{h,1}^{(1)}$};
\node at (7,-2) {(b)};

\draw[black, very thin,dashed] (0,0)--(0,1);
\draw[black, very thin,dashed] (0,0)--(2,1);
\draw[black, very thin] (0,0)--(0,-1);
\draw[black, very thin] (0,0)--(2,-1);

\draw[black,fill] (0,0) circle [radius=1mm];

\draw[black,fill=white] (0,1) circle [radius=1mm];
\draw[black,fill=white] (2,1) circle [radius=1mm];

\draw[black,fill] (0,-1) circle [radius=1mm];
\draw[black,fill] (2,-1) circle [radius=1mm];

\node[left=0.2cm] at (0,0) {$m_{c,M}^{(0)}$};
\node[left=0.2cm] at (0,1) {$m_{c,1}^{(1)}$};
\node[right=0.2cm] at (2,1) {$m_{h,1}^{(1)}$};
\node at (1,-2) {(a)};

\draw[black, very thin] (8,-4)--(6,-3);
\draw[black, very thin] (8,-4)--(8,-3);
\draw[black, very thin, dashed] (8,-4)--(6,-5);
\draw[black, very thin,dashed] (8,-4)--(8,-5);

\draw[black,fill] (8,-4) circle [radius=1mm];
\draw[black,fill] (6,-3) circle [radius=1mm];
\draw[black,fill] (8,-3) circle [radius=1mm];

\draw[black,fill=white] (6,-5) circle [radius=1mm];
\draw[black,fill=white] (8,-5) circle [radius=1mm];

\node[right=0.2cm] at (8,-4) {$m_{h,1}^{(0)}$};
\node[left=0.2cm] at (6,-5) {$m_{c,M}^{(-1)}$};
\node[right=0.2cm] at (8,-5) {$m_{h,M}^{(-1)}$};
\node at (7,-6) {(d)};

\end{tikzpicture}
\caption{A deconstruction of Figure \ref{2D_helix-like_config}, illustrating the four masses which have connections that extend outside of the generating unit cell $\Omega^{(0)}$. Springs and masses attributed to neighbouring unit cells are depicted by dashed lines and hollow circles}
\label{BFconds}
\end{center}
\end{figure}
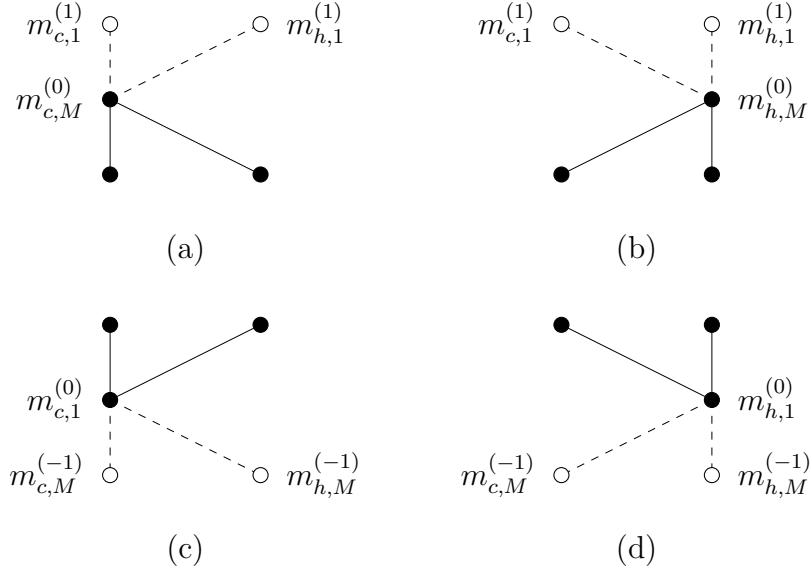

Representing the displacements of masses $m_{n,k}^{(j)}$ by $\textbf{u}_{n,k}^{(j)}, \, j \in \mathbb{Z}, k = 1,\dots,M, \,n = h,c$, we set the 
Bloch-Floquet condition as 
\begin{equation}
\textbf{u}^{(j)}_{n,k} = \text{e}^{\text{i}K jd}\textbf{u}^{(0)}_{n,k}, \quad  j \in \mathbb{Z},
\label{blochfloquetcond}
\end{equation}
where $d$ is the height of the unit cell, that is the pitch of the helix, and $K$ is the wavenumber acting solely in the $z$-direction. Thus, referring to Figure \ref{2D_helix-like_config}, we have
\[
\textbf{u}_{h,1}^{(1)} = \text{e}^{\text{i}K d}\textbf{u}_{h,1}^{(0)}, \quad
\textbf{u}_{h,M}^{(-1)} = \text{e}^{-\text{i}K d}\textbf{u}_{h,M}^{(0)}, \qquad
\textbf{u}_{c,1}^{(1)} = \text{e}^{\text{i}K d}\textbf{u}_{c,1}^{(0)}, \quad
\textbf{u}_{c,M}^{(-1)} = \text{e}^{-\text{i}K d}\textbf{u}_{c,M}^{(0)}.
\]
Viewing Figure \ref{2D_helix-like_config} with the Eigenglass app helps to demonstrate how the infinite structure may be built from a single unit cell, and uses red springs and masses to indicate where the Bloch-Floquet conditions are implemented (also see Figure \ref{BFconds}).
\\

To establish the governing equations, we begin with considering the masses on the helix that are  connected only to other masses within $\Omega^{(0)}$. 
In general, the linearised force $\textbf{F}_{i,j}$ between the $i^{th}$ and $j^{th}$ masses is given by the (responding force version of) Hooke's law

\[
\textbf{F}_{i,j} = - \kappa_{i,j} \, [ \textbf{e}_{i,j}^{(0)} \cdot (\textbf{u}_i - \textbf{u}_j) ] \,\textbf{e}_{i,j}^{(0)},
\]
where $\kappa_{i,j}$ is the stiffness of the spring between the two masses, $\textbf{e}_{i,j}^{(0)}$ is the unit vector in the direction from mass $i$ to $j$ at equilibrium, and $\textbf{u}_i$ and $\textbf{u}_j$ are small displacements of the respective masses. Therefore, the forces exerted on masses  $m^{(0)}_{h,k}$ by $m^{(0)}_{h,k\pm1}$ on the helix and on masses $m^{(0)}_{c,k}$ by $m^{(0)}_{c,k\pm1}$ on the central line are given by
\begin{equation}
    \textbf{F}_{(n,k),l} = -\kappa_{(n,k),l}\left[\frac{\textbf{p}_{n,l}-\textbf{p}_{n,k}}{|\textbf{p}_{n,l}-\textbf{p}_{n,k}|}\cdot(\textbf{u}_{n,k}-\textbf{u}_{n,l})\right]\frac{\textbf{p}_{n,l}-\textbf{p}_{n,k}}{|\textbf{p}_{n,l}-\textbf{p}_{n,k}|}, \quad k = 2,\dots,M-1,\, l = k\pm 1,
    \label{force1}
\end{equation}
for $n=h,c$.
For the cross interactions between the masses on the helix and the central line, still within the generating unit cell, we take into account both forces exerted on the mass $m_{h,k}$ by the masses $m_{c,k\pm 1}$: 
\begin{equation}
    \textbf{F}_{(h,k),(c,l)} = -\kappa_{(h,k),(c,l)}\left[\frac{\textbf{p}_{c,l}-\textbf{p}_{h,k}}{|\textbf{p}_{c,l}-\textbf{p}_{h,k}|}\cdot\left(\textbf{u}_{h,k}-\textbf{u}_{c,l}\right)\right]\frac{\textbf{p}_{c,l}-\textbf{p}_{h,k}}{|\textbf{p}_{c,l}-\textbf{p}_{h,k}|}, \quad k = 2,\dots,M-1,\, l = k\pm 1.
        \label{force2}
\end{equation}
Needless to say, the force exerted on the mass $m_{c,k}$ by $m_{h,k\pm 1}$ is $\textbf{F}_{(c,k),(h,l)} = -\textbf{F}_{(h,k),(c,l)}$. In (\ref{force1}), (\ref{force2}), the position vectors for each mass on the helical rim are given by
\[
\textbf{p}_{h,k} = \left(\alpha \cos \frac{2\pi (k-1)}{M}, \alpha \sin \frac{2\pi (k-1)}{M}, \frac{d(k-1)}{M} \right), \quad k=1,\dots,M,
\]
where $\alpha$ is the radius of the helix, and, clearly, $\textbf{p}_{c,k} = \left(0,0,d(k-1)/M\right),\,k=1,\dots,M$. \\
\subsection{Gyrocore helix}
In addition to the forces associated with Hooke's law, we introduce a gyroscopic spinner attached to each of the central line masses and amend the governing equation to include their gyroscopic force. To differentiate this physically chiral structure from the solely geometrically chiral helix-like structure, we henceforth refer to it as the gyrocore helix. The equations of motion for the $2(M-2)$ masses, which have no connections outside $\Omega^{(0)}$, may thus be written as follows
\begingroup
\addtolength{\jot}{0.7em}
\begin{align}
    &\ddot{\textbf{u}}_{h,k} - \mathbfcal{F}_{(h,k),k+1} - \mathbfcal{F}_{(h,k),k-1} - \mathbfcal{F}_{(h,k),(c,k+1)} - \mathbfcal{F}_{(h,k),(c,k-1)} = \textbf{0}, \label{helix_gov_eqn}\\ 
    &\ddot{\textbf{u}}_{c,k} - \mathbfcal{F}_{(c,k),k+1} - \mathbfcal{F}_{(c,k),k-1} - \mathbfcal{F}_{(c,k),(h,k+1)} - \mathbfcal{F}_{(c,k),(h,k-1)} -\tilde{\mathcal{G}}\textbf{R}\dot{\textbf{u}}_{c,k} = \textbf{0},
    \label{central_line_gov_eqn}
\end{align}
\endgroup
for \(k = 2,\dots,M-1\),
where $\mathbfcal{F}_{(n,k),\cdot} = \mathbf{F}_{(n,k),\cdot}/m_{n,k}$, and here and hereafter, dot on top of the variable denotes the time derivative. The presence of the terms with gyricity $\tilde{\mathcal{G}}$ and  the matrix $\textbf{R}$ in equation (\ref{central_line_gov_eqn}) are the mathematical equivalent of attaching a series of gyroscopic spinners to the central line masses, with 
$\textbf{R}$ describing the in-plane vorticity effect induced by the gyroscope
\begin{equation}
\textbf{R} = 
    \begin{pmatrix}
    0 & -1 & 0\\ 1 & 0 & 0\\ 0 & 0 & 0
    \end{pmatrix}.
\end{equation}
As the helix presented here has a right-handedness, the attached gyroscopes may either spin in the same direction (counterclockwise) or opposite (clockwise). The matrix $\textbf{R}$ describes a vorticity effect in agreement with the chirality of the helix.
\\

The remaining four masses $m^{(0)}_{c,M}$, $m^{(0)}_{h,M}$, $m^{(0)}_{c,1}$ and $m^{(0)}_{h,1}$ inside the generating unit cell are subject to forces from masses which lie outside $\Omega^{(0)}$. Their governing equations are given as 

\begin{figure}
    \centering
\begin{tikzpicture}
\node[inner sep=0pt] (sparsity) at (0,0)
    {\includegraphics[scale=0.2]{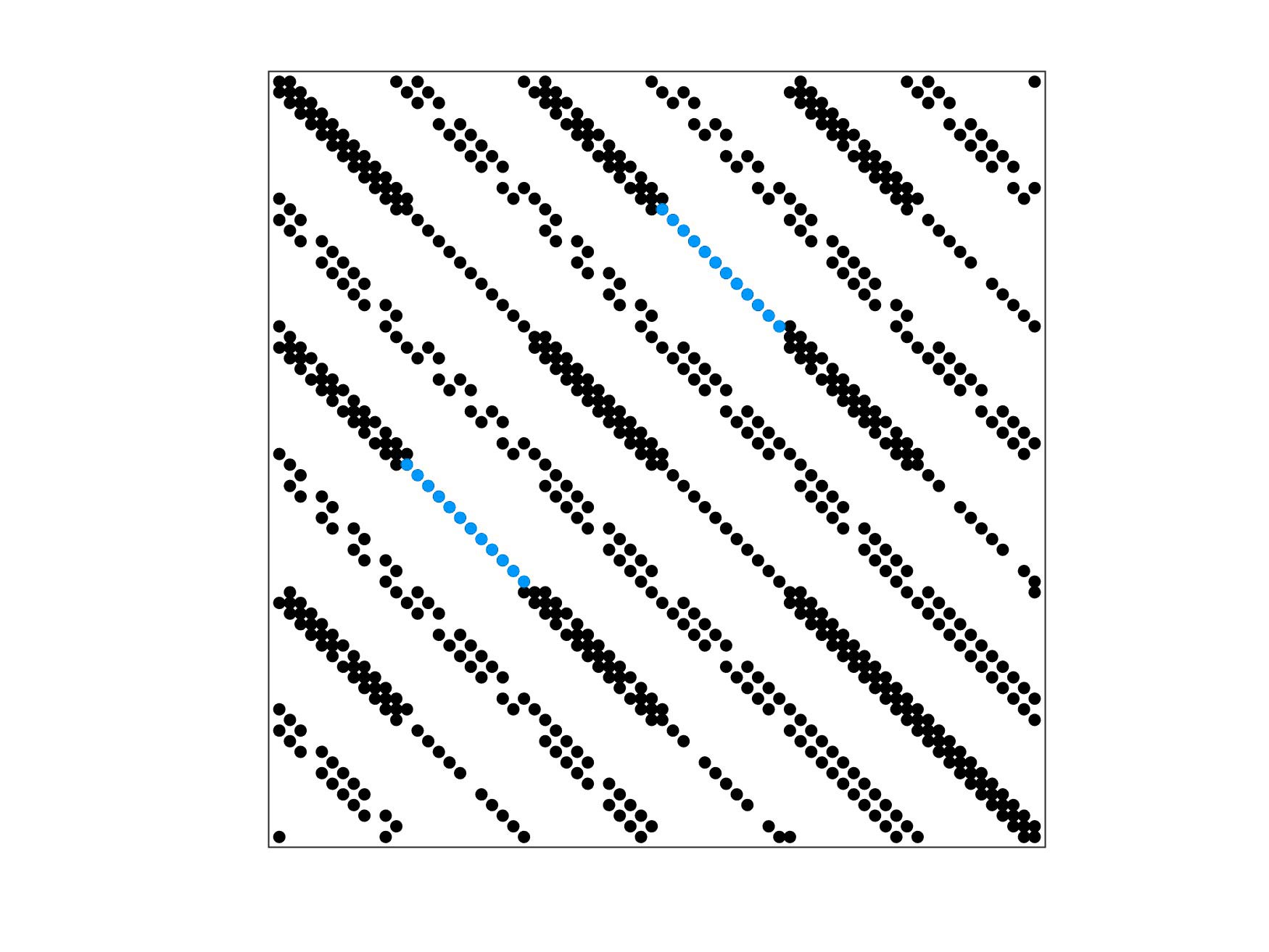}};
\node[black] at (-3.75,3.8) {$1$};
\node[black] at (-3.9,-3.4) {$72$};
\node[black] at (-3.45,4.2) {$1$};
\node[black] at (3.8,4.2) {$72$};
\end{tikzpicture}
    \caption{A sparsity plot, indicating the entries of the stiffness matrix $\mathbf{C}(K)$ (black) superimposed with the sparsity of the spinners matrix $\mathbf{A}$ (blue). There are clearly $6M=72$ algebraic equations for the choice of $M=12$.}
    \label{Asparsity}
\end{figure}
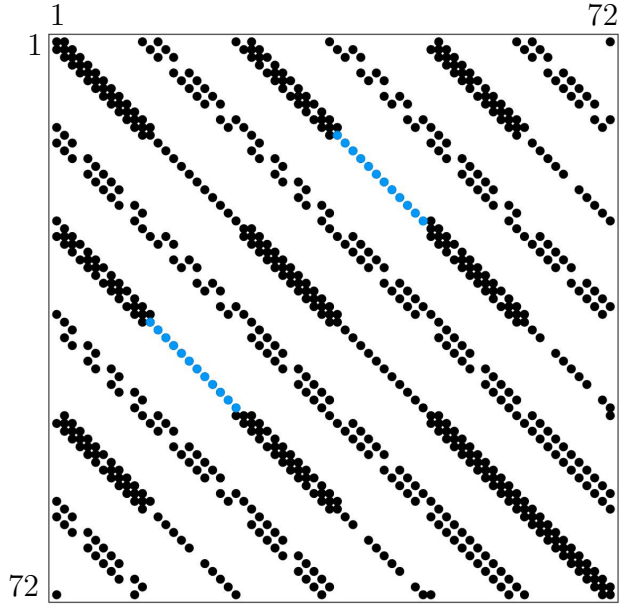

\begingroup
\addtolength{\jot}{0.7em}
\begin{align}
\begin{split}
    &\ddot{\textbf{u}}_{c,1} - \mathbfcal{F}_{(c,1),2} - \mathbfcal{F}_{(c,1),(h,2)} - \mathbfcal{F}^{(0,-1)}_{(c,1),M}  - \mathbfcal{F}^{(0,-1)}_{(c,1),(h,M)} - \tilde{\mathcal{G}}\textbf{R}\dot{\textbf{u}}_{c,1}= \textbf{0},\\
    &\ddot{\textbf{u}}_{c,M} - \mathbfcal{F}_{(c,M),M-1} - \mathbfcal{F}_{(c,M),(h,M-1)} - \mathbfcal{F}^{(0,1)}_{(c,M),1}  - \mathbfcal{F}^{(0,1)}_{(c,M),(h,1)} - \tilde{\mathcal{G}}\textbf{R}\dot{\textbf{u}}_{c,M}= \textbf{0},\\
    &\ddot{\textbf{u}}_{h,1} - \mathbfcal{F}_{(h,1),2} - \mathbfcal{F}_{(h,1),(c,2)} - \mathbfcal{F}^{(0,-1)}_{(h,1),M}  - \mathbfcal{F}^{(0,-1)}_{(h,1),(c,M)} = \textbf{0}, \\
    &\ddot{\textbf{u}}_{h,M} - \mathbfcal{F}_{(h,M),M-1} - \mathbfcal{F}_{(h,M),(c,M-1)} - \mathbfcal{F}^{(0,1)}_{(h,M),1}  - \mathbfcal{F}^{(0,1)}_{(h,M),(c,1)} = \textbf{0},
    \label{remaining_gov_eqns}
\end{split}
\end{align}
\endgroup
where the superscript now denotes the interaction between the generating unit cell $\Omega^{(0)}$ and the proceeding/preceding unit cells $\Omega^{(1)}$/$\Omega^{(-1)}$, with the forces given explicitly in the Supplementary Material (Section 1). Recall, that the displacements of masses outside the generating unit cell may be rewritten in terms of $u^{(0)}_{n,k}$ using the Bloch-Floquet condition in equation (\ref{blochfloquetcond}) and once again, $\mathbfcal{F}$ denotes relevant forces normalised by the associated masses.
Assuming that the solutions to equations (\ref{helix_gov_eqn}), (\ref{central_line_gov_eqn}) and (\ref{remaining_gov_eqns}) are time-harmonic, we let
\begin{equation}
\textbf{u}_{n,k} = \mathcal{U}_{n,k} \text{e}^{\text{i}\omega t},\quad k = 1,\dots,M, \,n = h,c,
\label{eigenfunctions}
\end{equation}
where \(\omega\) is the angular frequency, \(t\) is time and $\mathcal{U}_{n,k}$ are the displacement magnitudes. Upon substitution, (\ref{helix_gov_eqn}) and (\ref{central_line_gov_eqn}) may be reduced to a homogeneous system of linear equations given as 
\begingroup
\addtolength{\jot}{0.7em}
\begin{align*}
    &-\omega^2\mathcal{U}_{h,k} - \mathbfcal{F}_{(h,k),k+1} - \mathbfcal{F}_{(h,k),k-1} - \mathbfcal{F}_{(h,k),(c,k+1)} - \mathbfcal{F}_{(h,k),(c,k-1)} = \textbf{0}, \\ 
    &-\omega^2\mathcal{U}_{c,k} - \mathbfcal{F}_{(c,k),k+1} - \mathbfcal{F}_{(c,k),k-1} - \mathbfcal{F}_{(c,k),(h,k+1)} - \mathbfcal{F}_{(c,k),(h,k-1)} -i\tilde{\mathcal{G}}\omega\textbf{R}\mathcal{U}_{c,k} = \textbf{0},
\end{align*}
\endgroup
for \(k = 2,\dots,M-1\), with (\ref{remaining_gov_eqns}) deducible. Here we choose $\tilde{\mathcal{G}} = \gamma\omega/m_c$, defining a so-called active system, whereby the gyricity is proportional to the angular frequency of the mass displacements. For active systems, $\gamma$ denotes the spinner constant, which was was introduced in \cite{brun2012vortex}. 
Collecting the terms associated with linear-elastic forces, the homogeneous system of linear equations may be rewritten in matrix form as
\begin{equation}
   \left [\mathbf{C}(K)-\omega^2\left(\mathbf{I}-\mathbf{A}\right)\right ]\mathbf{U} = 0,
   \label{matrix_eqn}
\end{equation}
where $\mathbf{C}(K)$ and $\mathbf{A}$ are the normalised stiffness and spinners matrices, respectively, $\mathbf{I}$ is the identity matrix and $\mathbf{U}$ is the vector containing the three components of displacement for all $2M$ masses. Note that $\mathbf{C}(K)$ is a real matrix function periodic in $K$ and $\mathbf{A}$ is a Hermitian matrix, and in Figure \ref{Asparsity} we visualise its sparsity in relation to the stiffness matrix for a unit cell with $M=12$ masses. There are clearly $6M=72$ algebraic equations and recall that only the central line masses experience the gyroscopic force, meaning, in this instance, only 24 equations include this term.

\section{Solution}
\label{solutionsec}
\subsection{ Dispersion relation}
For non-trivial solutions of equation (\ref{matrix_eqn}), the following equation must hold
\begin{equation*}
    \text{det}\left [\mathbf{C}(K)-\omega^2\left(\mathbf{I}-\mathbf{A}\right)\right ] =0.
\end{equation*}
By defining a normalised Brillouin zone $|Kd|\leq \pi$, and subsequently fixing the value of  $K$, the above dispersion equation may be solved numerically for the eigenfrequencies $\omega$. Substitution of these eigenvalues into equation (\ref{matrix_eqn}) gives the eigenvectors $\textbf{U}$ which, by taking the real part of equation (\ref{eigenfunctions}), define the eigenfields of the system. Through the numerical assembly of dispersion diagrams, the relationship between the angular frequency \(\omega\) and wavenumber \(K\) may be illustrated and analysed. Although the appearance of the dispersion diagram will change significantly depending on the tuning of mass and stiffness, there are always four sets of branches present, whose associated eigenmodes correspond to radial or longitudinal vibrations, either on the helix or central line. \\

\subsection{Low-frequency eigenfields}

\begin{figure}[h!]
\centering
\subfloat{%
\resizebox*{9cm}{!}{\includegraphics{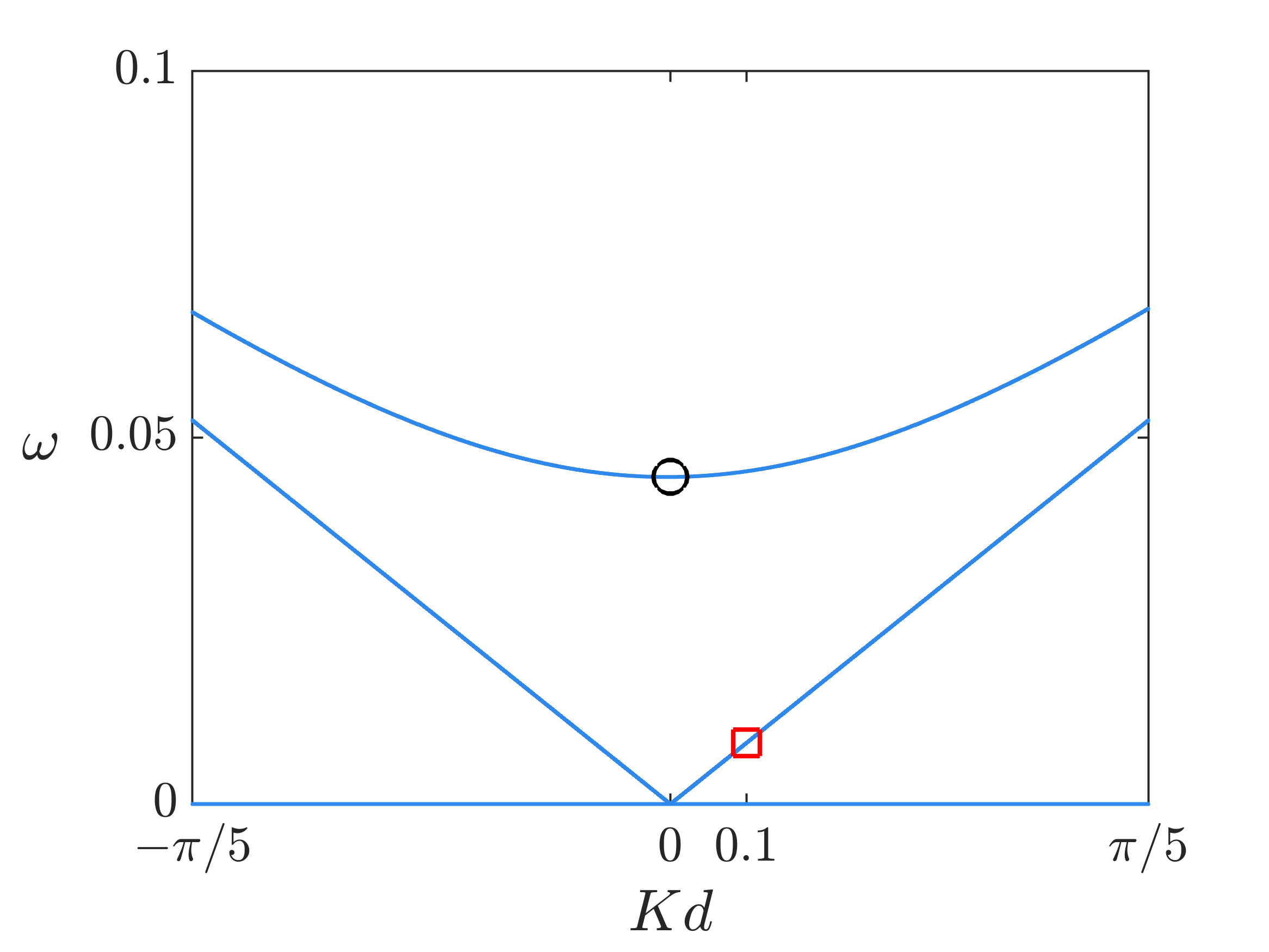}}}
\caption{A focused, low frequency section of the dispersion diagram for a gyrocore helix indicating the smallest non-zero frequency (black circle marker) and the acoustic branch (red square marker).}
\label{lowfrequencydisp}
\end{figure}

\begin{figure}[h!]
    \centering
    \begin{subfigure}{0.45\textwidth}
  \includegraphics[width=\textwidth]{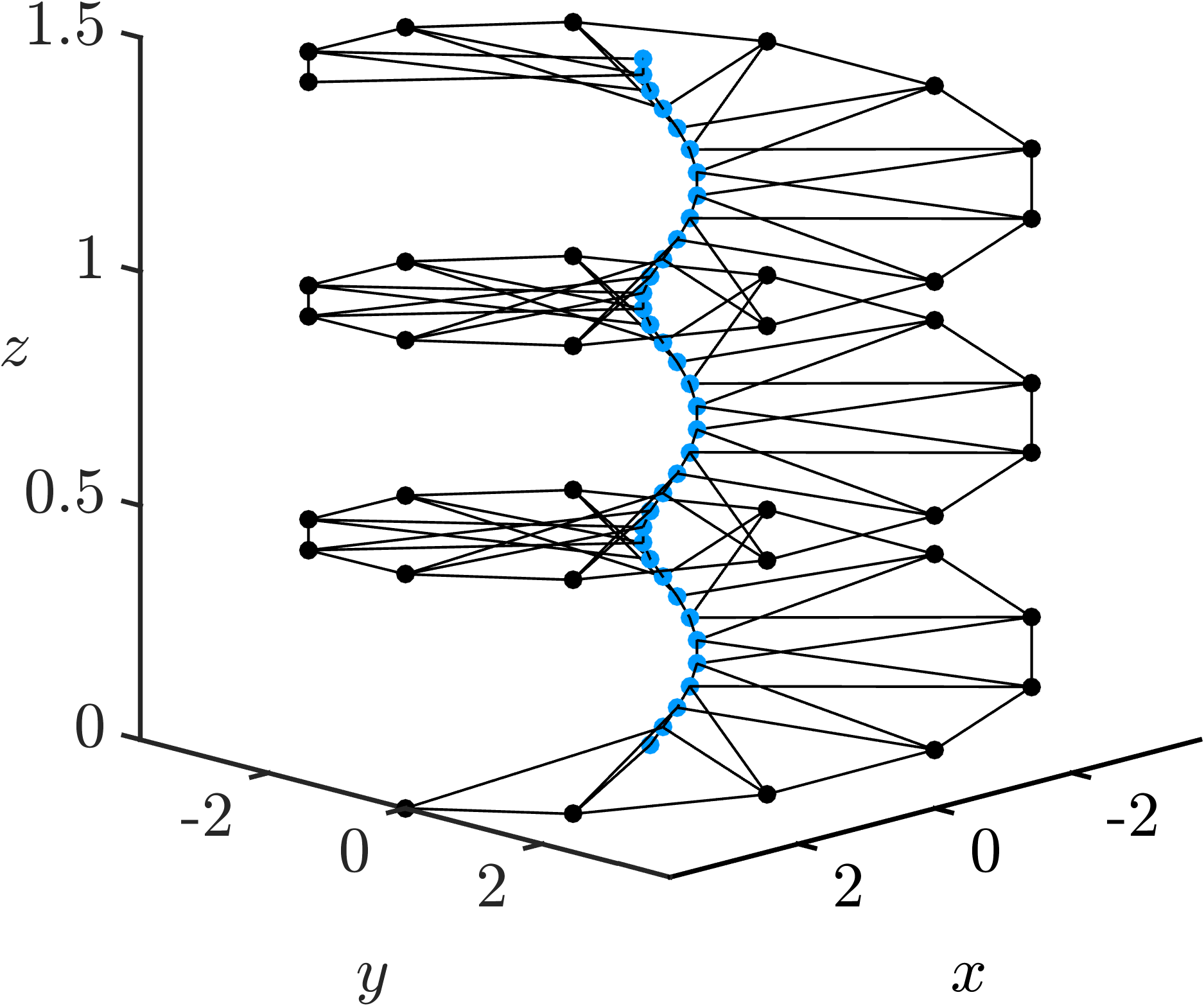}
  \caption{}
  \end{subfigure} \qquad\begin{subfigure}{0.45\textwidth}
  \includegraphics[width=\textwidth]{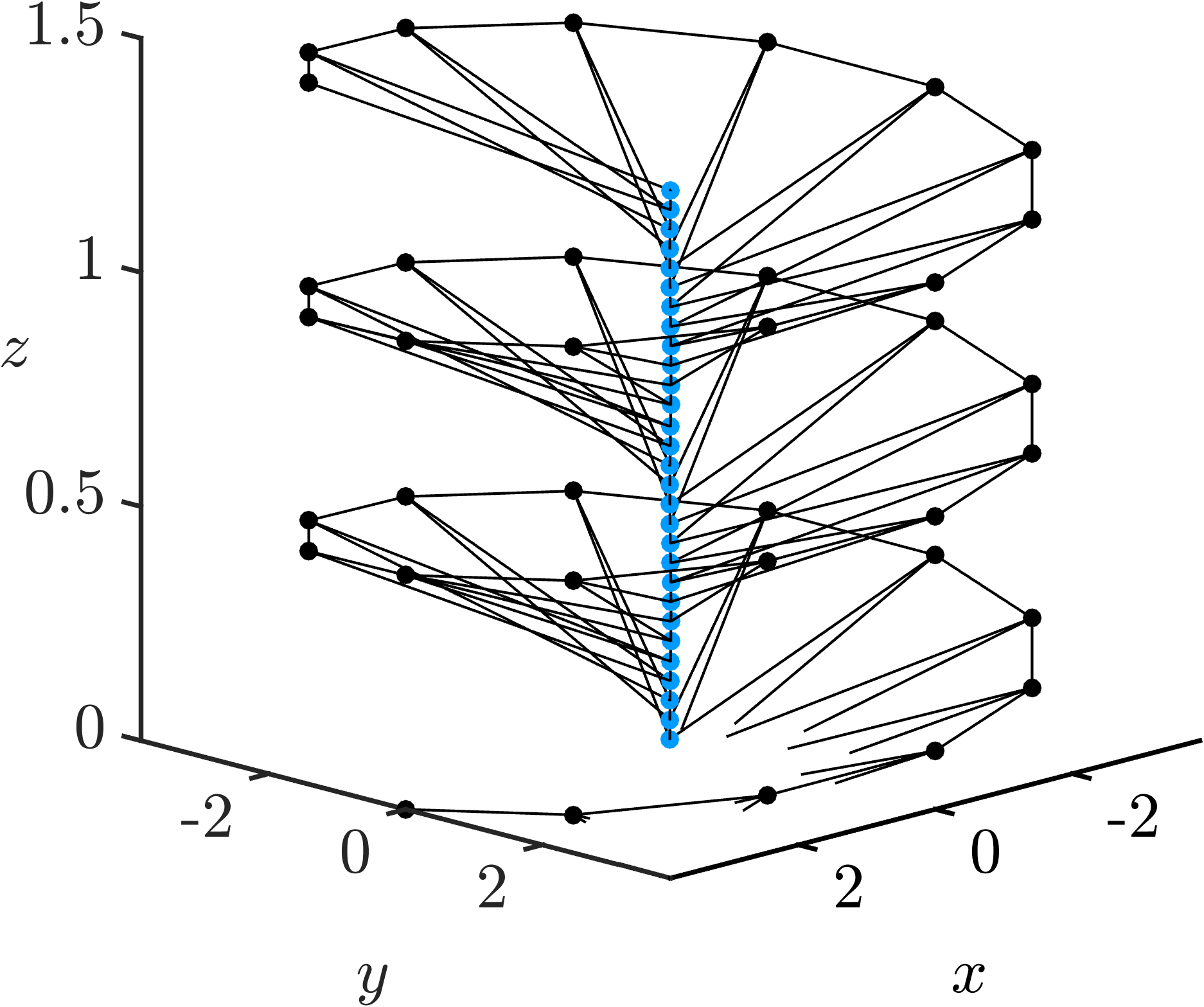}  
  \caption{}
  \end{subfigure} 
\caption{The low frequency transverse mode (a) and the `longitudinal rod' vibrational mode (b) for the gyrocore helix, identified by the black circle and red square markers in Figure \ref{lowfrequencydisp}, respectively. \mychar{}}
\label{lowfrequencymodes}
\end{figure}

As an illustration, we locate $M =12$ masses on both the helix and central line, impose a small pitch height of $d=0.5$, a radius of $\alpha = 4$ and $\gamma = -10$. We tune the structure to elicit specific vibrational behaviour by stiffening the radial and central springs, and making the central line of masses significantly heavier than the ones on the helix. In Figure \ref{lowfrequencydisp}, we focus on the low frequency behaviour of the gyrocore helix. 
The smallest non-zero eigenfrequency, identified with the black circle marker, is associated with transverse displacements of the central line masses with zero group velocity. Since the gyroscopes are, of course, unable to perform straight line translations, the masses for this mode follow a curved trajectory - a subtlety that can easily be understood utilising the Eigenglass visualiser. Although the associated dispersion branch appears to be independent of gyricity (since the symmetry of Figure \ref{lowfrequencydisp} remains intact)  there is in fact a negligible increase in $\omega$ for a corresponding change in $\gamma$.
Stemming from the origin, the acoustic branch exhibits finite effective refractive index, which we note is unaffected by the gyricity of the system. The acoustic branch also represents a special mode corresponding to a uniform, longitudinal translation of the central line masses (see Figure \ref{lowfrequencymodes}(b)). Small values of $Kd$, (see square red marker in Figure \ref{lowfrequencydisp}), create the illusion of a long rigid rod oscillating up and down.
\\

\section{Results}
\label{Results}

\subsection{Splitting of eigenfrequencies}

\begin{figure}[h!]
\includegraphics[scale=0.55]{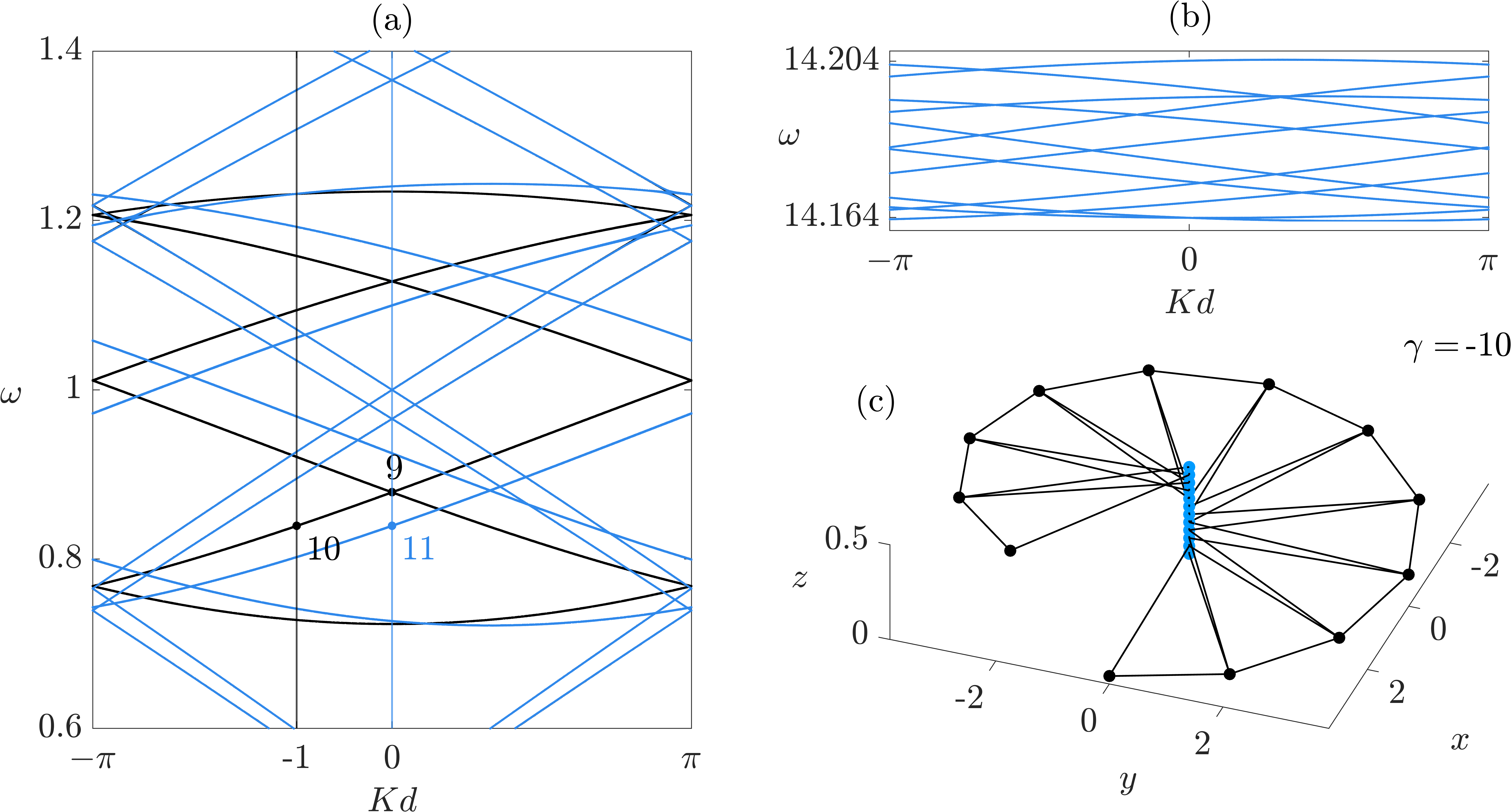}
    \caption{(a) A focused section of the dispersion diagram for the gyrocore helix for $\gamma=-10$. Black dispersion branches have been superimposed to indicate the position of the dispersion curves before the introduction of the gyroscopes. The eigenfields corresponding to the points marked by markers 9, 10 and 11 are shown in the Figures 9, 10 and 11, respectively. (b) Part of the same dispersion diagram given in (a), focusing on the higher-frequency branches associated with radial modes on the helix. (c) The generating unit cell of the infinite gyrocore helix. Using the Eigenglass app, one can see how the dispersion curves translate across the diagram as the magnitude of the spinner constant $\gamma$ is increased. \mychar{}}
    \label{gyrodispersion}
\end{figure}

Under the same tuning as presented in Section \ref{solutionsec}, we demonstrate that the gyrocore helix breaks the symmetry of the associated dispersion diagram, splitting the paired eigenfrequencies along $Kd=0$. 
In Figure \ref{gyrodispersion}(a), it is apparent that the addition of clockwise-spinning gyroscopes produces a shift of the dispersion curves to the right, specifically the ones describing radial shearing modes on the central line. To emphasise this shift, dispersion curves obtained from a helix-like structure with no gyroscopes (which are of course symmetric about $Kd = 0$) are superimposed in black. Also referring to the same figure, we note that the branches associated with longitudinal modes on the helix and central line are unaffected by the addition of the gyroscopes. However a shifting, in addition to a degree of skewing, also occurs for the higher frequency branches associated with radial modes on the helix, depicted in Figure \ref{gyrodispersion}(b). In fact, the amount of shift we observe is proportional to the gyricity $\tilde{\mathcal{G}}$ through the value of the spinner constant $\gamma$. Interestingly, a leftward shift of the dispersion curves may be visible if the vorticity of the gyroscopes is reversed 
and they spin in agreement with the geometric chirality of the helix. 
\subsection{Group velocity}
Approximating the dispersion branches as straight lines, the split eigenfrequencies along $Kd=0$ may be determined from the eigenfrequency $\omega_0$, that was originally paired before the introduction of the gyroscopes, and the group velocity $v_g=\text{d}\omega/\text{d}K$ as 
\begin{equation*}
    \omega^{\pm} -\omega_0 = \pm \mathcal{G}v_g. \label{Faraday_equiv_formula}
\end{equation*}
Here, $\omega^{\pm}$ are the higher and lower split eigenfrequency values, respectively, and $\mathcal{G}(\tilde{\mathcal{G}},m_c) = \gamma/10$. Indeed, for a fixed frequency, it is clear that the direction of the group velocity is dependent on the vorticity of the gyroscopic spinners
\begin{equation*}
    v_g = \pm\frac{\omega^{\pm}-\omega_0}{\mathcal{G}},
    \label{groupvel2}
\end{equation*}
which, after using the substitution $\omega^{\pm}_{*} = \omega^{\pm}-\omega_0$, gives equation (\ref{groupvelocity}). It is apparent then, that the interplay between these two chiralities acts as a mechanical analogue of electromagnetic induction, whereby the direction of current is determined by the polarisation of the magnetic field in relation to the handedness of the coil. \\
\subsection{
Unidirectional waveforms in a helicolidal gyroscopic waveguide}

Applying Bloch-Floquet conditions at the macrocell's boundary introduces spatial periodicity into the helix-like structure. At $Kd=0$, a distinct point emerges in the Brillouin zone, where the wavevector corresponds to a spatially invariant wave (a standing mode), formed by the interference of counter-propagating waves. After injecting a physical chirality along the central line, we observe a shifting of the dispersion branches associated with radial shearing modes. Each split eigenfrequency will correspond to an eigenmode with either positive or negative group velocity. A natural inquiry arises: How do these non-zero group velocity eigenmodes at $Kd=0$ compare with those before the shift, which share identical frequency but different wavenumber? In essence, we would like to address the question: If the eigenfrequency of a gyrocore helix at $Kd=0$ coincides with the helix-like structure for, say, $Kd = -1$, would there also be correspondence between their eigenmodes?
\\
\\
\begin{figure}[h!]
    \centering
    \includegraphics[scale=0.5]{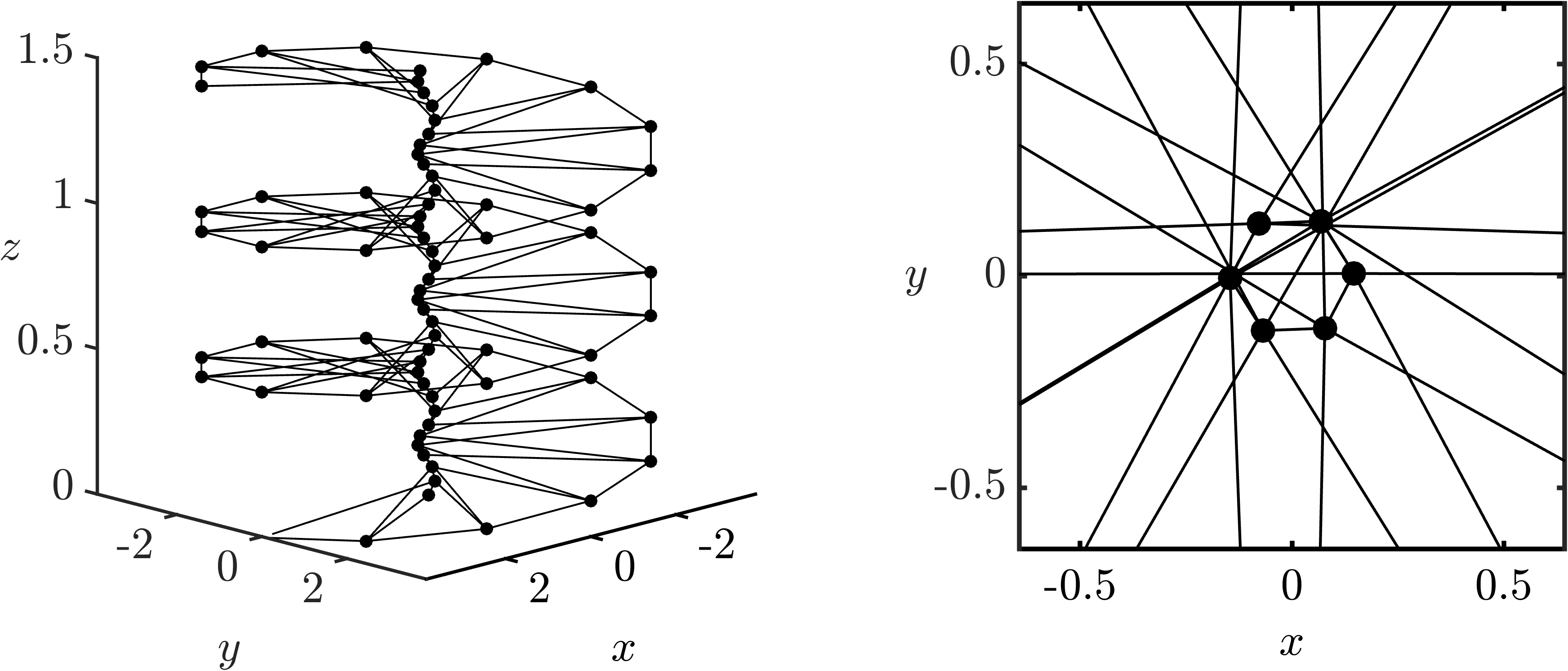}
    \caption{A lateral view of three cells of the helix-like structure and a bird's-eye view focused only on the shear radial displacements of the central line masses for the mode indicated by the marker labelled 9 in Figure \ref{gyrodispersion}(a). \mychar{}}
    \label{gyro_mode_stand}
\end{figure}
\\
\begin{figure}[h!]
    \centering
        \begin{tikzpicture}
\node[inner sep=0pt] (gyro mode standing) at (0,0)
  {\includegraphics[scale=0.5]{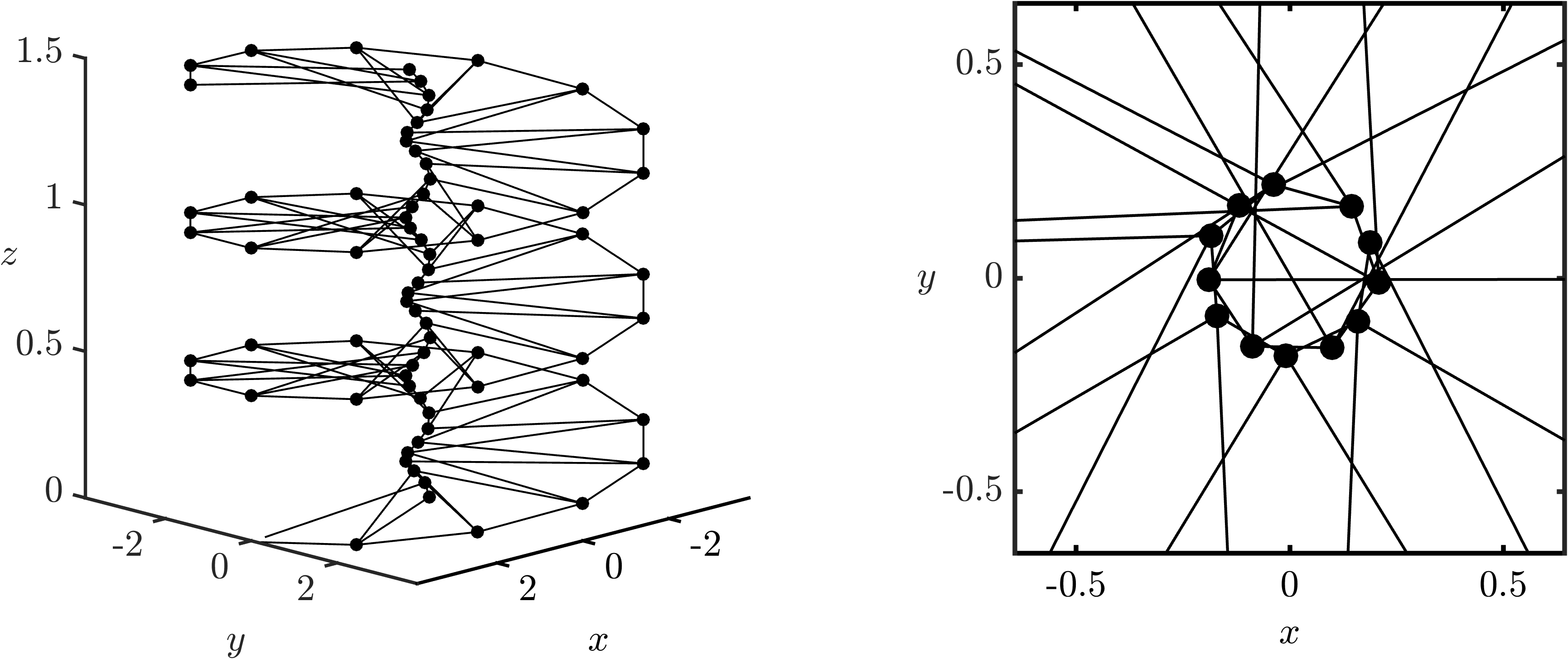}};
  \draw[-to, very thick] (0.2,0) to (0.2,1);
  \node[black, left] at (0.2,0.5) {$v_g$};
  \node[black, left] at (-0.6,4.2) {rotation of the central line};
  \draw [-to, very thick, rotate=-90] (-3.2,-3.2) arc [start angle=-340, end angle=-30, x radius=0.25cm, y radius=0.5cm];
    \end{tikzpicture}
    \caption{A lateral view of three cells of the helix-like structure and a bird's-eye view focused only on the shear radial displacements of the central line masses for the mode indicated by the marker labelled 10 in Figure \ref{gyrodispersion}(a). \mychar{}}
    \label{gyro_mode_disp}
\end{figure}
\\
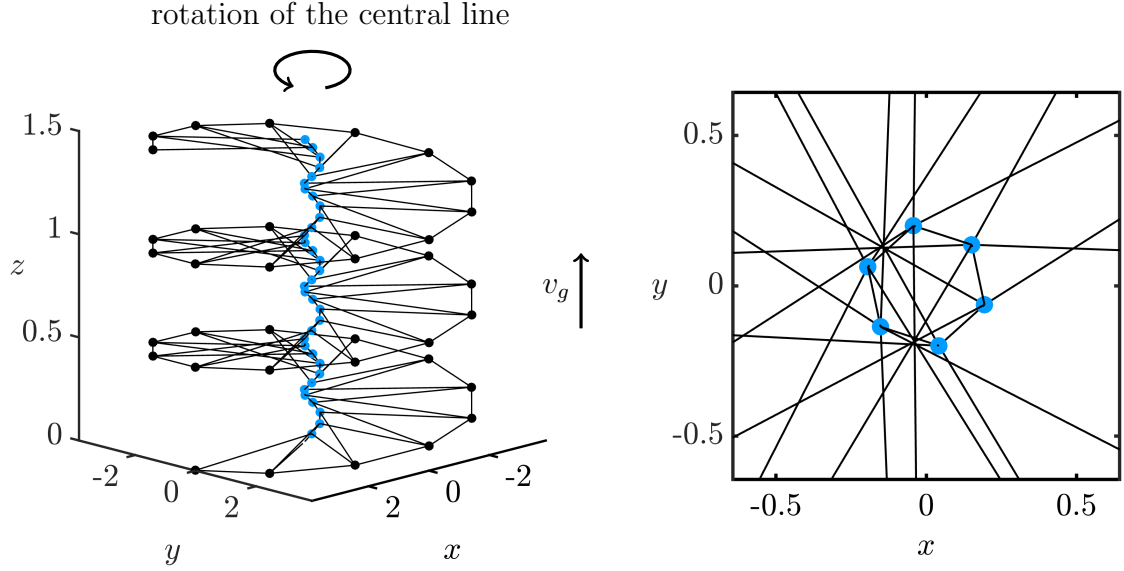
\begin{figure}[h!]
\centering
        \begin{tikzpicture}
\node[inner sep=0pt] (gyro mode standing) at (0,0)
  {\includegraphics[scale=0.5]{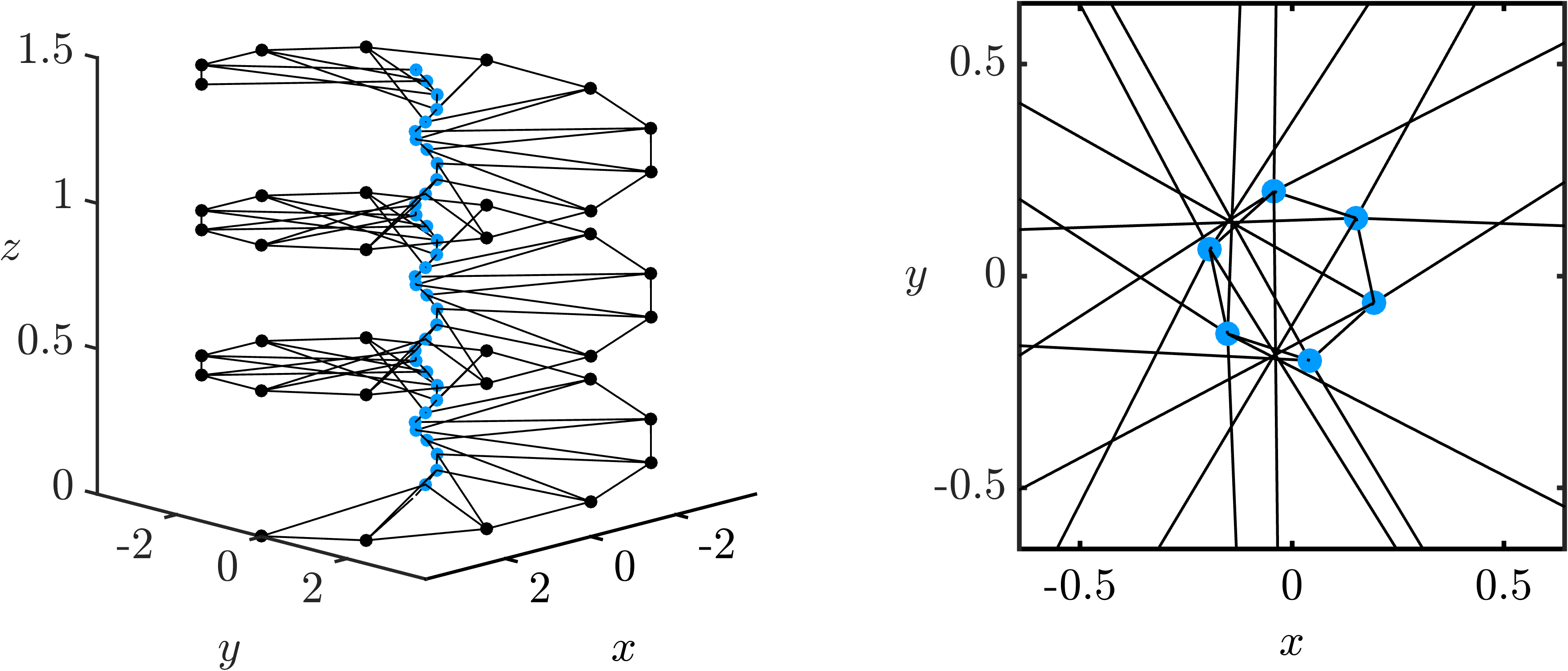}};
  \draw[-to, very thick] (0.2,0) to (0.2,1);
  \node[black, left] at (0.2,0.5) {$v_g$};
  \node[black, left] at (-0.6,4.2) {rotation of the central line};
  \draw [-to, very thick, rotate=-90] (-3.2,-3.2) arc [start angle=-340, end angle=-30, x radius=0.25cm, y radius=0.5cm];
    \end{tikzpicture}
\caption{A lateral view of three cells of the gyrocore helix and a bird's-eye view focused only on the shear radial displacements of the central line masses for the mode indicated by the marker labelled 11 in Figure \ref{gyrodispersion}(a). \mychar{}}
    \label{gyro_mode_gyro}
\end{figure}

\noindent In Figures \ref{gyro_mode_stand}, \ref{gyro_mode_disp} and \ref{gyro_mode_gyro}, we plot a lateral and bird's-eye view of the structures, with a close-up on the central line of masses, for the following cases (see markers in Figure \ref{gyrodispersion}(a)):
\begin{enumerate}

\item[Figure 9:]  one of the paired standing modes at $Kd=0$ for the helix-like structure with zero group velocity,

\item[Figure 10:] the eigenmode corresponding to the lower-frequency branch at $Kd=-1$ for the helix-like structure,

\item[Figure 11:] the eigenmode with an identical eigenfrequency and group velocity as above, obtained at $Kd=0$ for the gyrocore helix.  
\end{enumerate}

We choose $Kd=-1$ for the helix-like structure in order to target the identical eigenfrequency which lies on the line $Kd=0$ in Figure \ref{gyrodispersion}(a). The base case displayed in Figure \ref{gyro_mode_stand} is clearly a standing mode, formed by the interference of an upward and a downward travelling wave. From the bird's-eye view, the 12 central line masses (within each macrocell) overlap to form a hexagon, which appears to oscillate in size as the shear displacements fluctuate. Introducing negative phase velocity into the system creates latency between the central line masses (from top to bottom). From above, this latency means the central masses no longer overlap and the central polygon appears to rotate in place, as evident in Figure \ref{gyro_mode_disp}. Finally, in Figure \ref{gyro_mode_gyro}, we target the same frequency on the dispersion branch but at $Kd=0$ for the gyrocore helix, made possible by the shift. Indeed, this shear radial eigenmode of the gyrocore helix closely resembles that of an identical structure without a physical chirality. Despite a difference in wavenumber, an upward travelling wave packet is present in both Figures \ref{gyro_mode_disp} and \ref{gyro_mode_gyro} due to the positive group velocity, and both eigenmodes exhibit an internal helical structuring of the central line with identical direction of spin and handedness. The notable difference with the gyrocore helix however, is a net energy transfer within the wave packet, despite an absence of spatial variation across the macrocell.
\\

\section{Conclusion}
\label{Conclusion}
In this paper, we establish a method to induce unidirectional wave propagation in a helicoidal waveguide through considering a combination of geometric and physical chiralities.
The introduction of the gyroscopic force to the governing equations prompts a breaking of symmetry of the associated dispersion diagram, the direction of shift being dependent on the vorticity of the gyroscopic spinners, ultimately establishing a connection between gyricity and the group velocity direction for a given frequency. The gyroscopes, along with the subsequent coupling between components of the displacement vector, gives rise to spatially invariant modes with non-zero group velocity. These modes, exemplified by the gyrocore helix at $Kd=0$, represent a unique class of waveforms, where energy propagates through the system, either up or down, depending on the gyricity of the gyrocore. We note that the effect of preferential direction disappears if the gyricity of the central line is zero. This ability to dynamically tune the group velocity direction presents possibilities for the design of novel waveguides and energy harvesting devices, where controlling wave propagation at specific frequencies and directions is essential. 


\section{Acknowledgements}
F. J. P. Allison gratefully acknowledges the Engineering and Physical Sciences Research Council [grant number EP/V52007X/1] for funding his studentship.

\bibliography{bibliography.bib}
\end{document}